\documentclass[aip,
amsmath,amssymb,
reprint,
]{revtex4-1}
\usepackage{graphicx}
\usepackage{dcolumn}
\usepackage{bm}
\usepackage{siunitx}
\usepackage[utf8]{inputenc}
\usepackage[T1]{fontenc}
\usepackage{mathptmx}
\usepackage{etoolbox}
\usepackage{overpic}
\usepackage{pict2e}
\usepackage[wby]{callouts}
\usepackage[hidelinks]{hyperref}
\usepackage{multirow}
\usepackage{setspace}
\usepackage[english]{babel}
\usepackage{lineno}
\usepackage[version=4]{mhchem}
\usepackage{tabularx}
\usepackage{booktabs}
\usepackage{array}

\makeatletter
\def\@email#1#2{%
 \endgroup
 \patchcmd{\titleblock@produce}
  {\frontmatter@RRAPformat}
  {\frontmatter@RRAPformat{\produce@RRAP{*#1\href{mailto:#2}{#2}}}\frontmatter@RRAPformat}
  {}{}
}
\makeatother
\begin{document}

\preprint{AIP/123-QED}
\title[]{Room temperature reactive sputtering deposition of titanium nitride with high sheet kinetic inductance}
\author{Juliang Li}
\email{juliangli@uchicago.edu}
\affiliation{Argonne National Laboratory, 9700 South Cass Ave., Lemont, IL, 60439, USA}
\affiliation{University of Chicago, 5640 South Ellis Ave., Chicago, IL, 60637, USA}
\author{Peter S Barry}
\affiliation{School of Physics and Astronomy, Cardiff University, Cardiff, CF24 3AA, Wales, UK}
\author{Tom Cecil}
\affiliation{Argonne National Laboratory, 9700 South Cass Ave., Lemont, IL, 60439, USA}
\author{Marharyta~Lisovenko}
\affiliation{Argonne National Laboratory, 9700 South Cass Ave., Lemont, IL, 60439, USA}
\author{Volodymyr~Yefremenko}
\affiliation{Argonne National Laboratory, 9700 South Cass Ave., Lemont, IL, 60439, USA}
\author{Gensheng~Wang}
\affiliation{Argonne National Laboratory, 9700 South Cass Ave., Lemont, IL, 60439, USA}
\author{Serhii~Kruhlov}
\affiliation{\parbox[t]{0.8\textwidth}{Department of Physics, Drexel Univeristy,  3141 Chestnut St., Philadelphia, PA 19104, USA}}
\author{Goran~Karapetrov}
\affiliation{\parbox[t]{0.8\textwidth}{Department of Physics, Drexel Univeristy,  3141 Chestnut St., Philadelphia, PA 19104, USA}}

\author{Clarence Chang}
\affiliation{Argonne National Laboratory, 9700 South Cass Ave., Lemont, IL, 60439, USA}
\affiliation{University of Chicago, 5640 South Ellis Ave., Chicago, IL, 60637, USA}
\affiliation{\parbox[t]{0.8\textwidth}{Kavli Institute for Cosmological Physics, U. Chicago, 5640 South Ellis Ave., Chicago, IL, 60637, USA}}

\begin{abstract}
Superconducting thin films with high intrinsic kinetic inductance $L_{k}$ are important for high-sensitivity detectors, enabling  strong coupling in hybrid quantum systems, and enhancing nonlinearities in quantum devices. We report the room-temperature reactive  sputtering of titanium nitride thin films with a critical temperature $T_{c}$ of \SI{3.8}{K} and a thickness of \SI{27}{nm}.  Fabricated into resonators, these films exhibit a sheet kinetic inductance $L_{k, \square}$ of 394~$\textrm{pH}/\square$, as inferred from resonant frequency measurements. 
X-ray diffraction analysis confirms the formation of stoichiometric TiN, with no residual unreacted titanium. The films also demonstrate a characteristic sheet resistivity of 475~$\Omega/\square$, yielding an impedance an order of magnitude higher than conventional 50~$\Omega$ resonators. This property could enhance microwave single\textendash photon coupling strength by an order of magnitude, offering transformative potential for hybrid quantum systems and quantum sensing. Furthermore, the high $L_{k}$ enables Kerr nonlinearities comparable to state\textendash of\textendash the\textendash art quantum devices. Combined with its relatively high $T_{c}$, this thin film presents a promising platform for superconducting devices, including  amplifiers and qubits operating at higher temperatures. 
\end{abstract}
\maketitle

\section{Introduction}
Disordered superconducting materials with high kinetic inductance have opened new opportunities in quantum electronics and quantum sensing. Low energy dissipation combined with high overall inductance in nanoscale-size quantum devices has enabled the design of high fidelity superconducting circuits such as microwave kinetic inductance detectors (mKIDs)~\cite{day2003,bertrand2021,henry2010} and circuit QED experiments~\cite{samkharadze2016, maleeva2018}. Fine tunability of the kinetic inductance via DC current bias has extended further the application of these quasi-amorphous materials to novel devices such as frequency tunable superconducting resonators\cite{li2024, xu2019apl}, superconducting phase shifters~\cite{che2017,thakur2020}, ultra-sensitive current sensors\cite{kher2016}, and near quantum-limited parametric amplifiers\cite{farzad2024,malnou2021, zapata2024}. Large zero-point fluctuations of the electric field in these materials offers increased spin-photon coupling that could extend significantly the fidelity of quantum bits.~\cite{samkharadze2016,landig2018,samkharadze2018} The same disorder that leads to high kinetic inductance was shown to drive up the quantum coherence resulting in high kinetic inductance qubits\cite{faramrzi2021, winkel2020, grunhaupt2019, hazard2019}, qubit protection devices\cite{kerman2010,bell2014,dempster2014}, and mm-wave quantum devices.~\cite{anferov2020}

Disordered superconducting nitrides' (e.g. NbN, TiN, NbTiN) resilience to external high magnetic fields~\cite{xu2023, wyatt2023,frasca2024} makes them suitable for spin-photon coupling applications in which external magnetic field is required to excite spin modes. Other desirable properties of these materials are fabrication compatibility - for instance room temperature deposition that is compatible with liftoff processing and low melting point metals already present on the wafer surface - and relatively high superconducting critical temperature. But overall sheet kinetic inductance has been relatively low, compared to materials like granular aluminum~\cite{, grunhaupt2018}, necessitating high aspect ratio nanowires which impose stringent tolerances to geometrical imperfections.~\cite{samkharadze2016,niepce2019,pitavidal2020,maurand2021}
Recently, high quality TiN films with large kinetic inductance have been grown using an ALD process that requires deposition temperature $\approx 300^{\circ}$~C and hours long deposition times.~\cite{shearrow2018,coumou2013} The high process temperatures are often incompatible with the overall thermal budget imposed by the downstream device fabrication process.

In this work, we report our initial results on room-temperature reactive sputter-deposited TiN thin films, which exhibit unusually high kinetic inductance, high normal-state sheet resistance, high superconducting critical temperature, and a sharp superconducting transition. These properties may be attributed to short-range disorder, as suggested earlier\cite{ohya2013,driessen2012}. Further investigation into the effects of disorder will be a continued focus of this research project.

\section{resonator design and fabrication}
\label{sec::design}
To characterize the $L_{k, \square}$ of the sputtered TiN film we fabricated three groups of resonators and calculated the $L_{k,\square}$ values from their measured resonant frequency.
 All resonators were fabricated on the same $4''$ wafer. The fabrication started with growing our TiN films at room temperature using a reactive DC-magnetron  sputtering process with a high-purity titanium target. The deposition was carried out in an \ce{Ar}/\ce{N2} gas mixture with flow rates of \SI{30}{sccm} and \SI{3.6}{sccm}, respectively, at a fixed DC power of \SI{200}{W}. The working pressure during deposition was maintained at \SI{3}{mTorr}, and the resulting deposition rate was approximately \SI{9}{\angstrom}/minute (see Table~\ref{tab:paras} and Fig.~\ref{fig::chamber}). For resonator formation, we utilized reactive ion etch (RIE) with \ce{CHF3}+\ce{SF6} gas mixture. Using maskless direct laser writing lithography, we achieved an inductor width of \SI{0.7}{\micro\meter}. To ensure complete removal of the TiN, an overetch was applied, resulting in approximately \SI{200}{nm} etched into the underlying Si substrate, as confirmed by post-process profilometer measurements. Following the etching of the TiN, a layer of photoresist was spin-coated onto the wafer and subsequently patterned using a maskless aligner. Nb structure (\SI{120}{nm} thick) were shaped using the liftoff process. Immediately before Nb sputtering, a 90 seconds ion milling was done in the same vacuum chamber to ensure good galvanic contact between the TiN and Nb layers.

Figure \ref{fig::alphadesign} shows the first group of fabricated resonators which we have named the `alpha' design. Two classes of lumped element LC resonators were fabricated on the same readout line. One class has both the interdigital capacitors and meander inductors fabricated from a Nb film, and the other class has the identical physical geometry only with inductors fabricated from TiN. There are three resonators within each class with stepped inductor width of 18, 20 and \SI{22}{\micro\meter} while all other parameters are kept the same. The difference in resonant frequency between the two classes is a direct result of the extra $L_{k,\square}$ in the TiN. The wide inductor design mediates the resonance difference between the two classes of resonators by reducing the participation ratio of the kinetic inductance $L_{k}$ in the total inductance, and all six resonances are in the bandwidth of one HEMT amplifier. The measured resonant frequency and internal quality factor are listed in Table \ref{tab:alphaparas}. The measured quality factors of $\sim10^{4}$ are likely limited by the low resistivity silicon wafer used in the fabrication process.
\begin{figure}
     \begin{overpic}[abs,unit=1pt,scale=.01, angle = -90, width=0.45\textwidth]{alphaall6.pdf}
     \put(10,80){\color{black}\large{(a)}}
     \put(40,20){\color{black}\large{(\SI{18}{\micro\meter})}}
     \put(125,20){\color{black}\large{(\SI{20}{\micro\meter})}}
     \put(170,20){\color{black}\large{(\SI{22}{\micro\meter})}}
     \put(100,70){\linethickness{0.25mm}\color{black}\polygon(2,0)(22,0)(22,44)(2,44)}
     \put(103,10){\linethickness{0.25mm}\color{black}\polygon(2,0)(22,0)(22,44)(2,44)}
     \put(52, -165){\color{white}\linethickness{0.25mm}
     \frame{\includegraphics[scale=.27, angle=0]{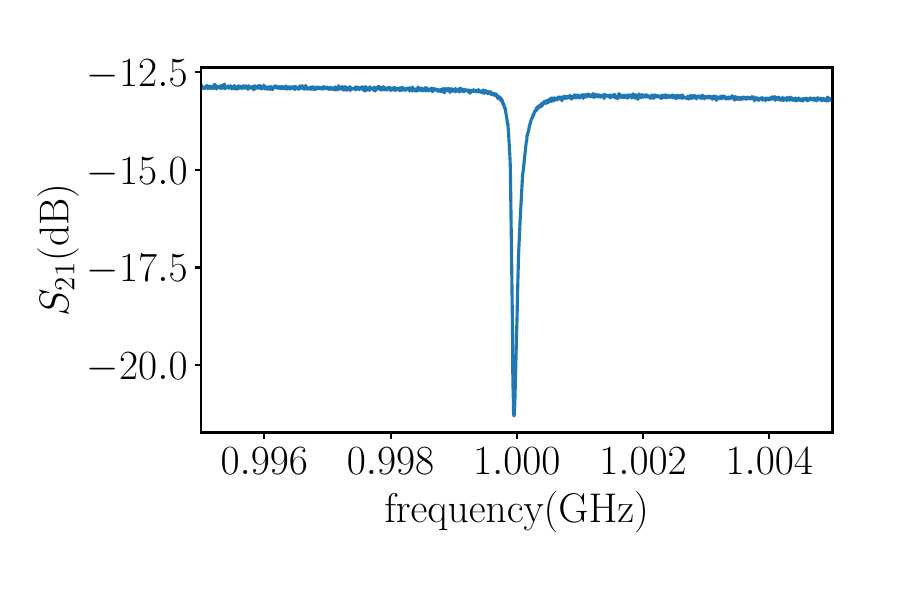}}}
     \put(82,-135){\color{black}\large{(e)}}
     \put(-8, -150){\color{black}\linethickness{0.25mm}
     \frame{\includegraphics[scale=.05, angle=-90]{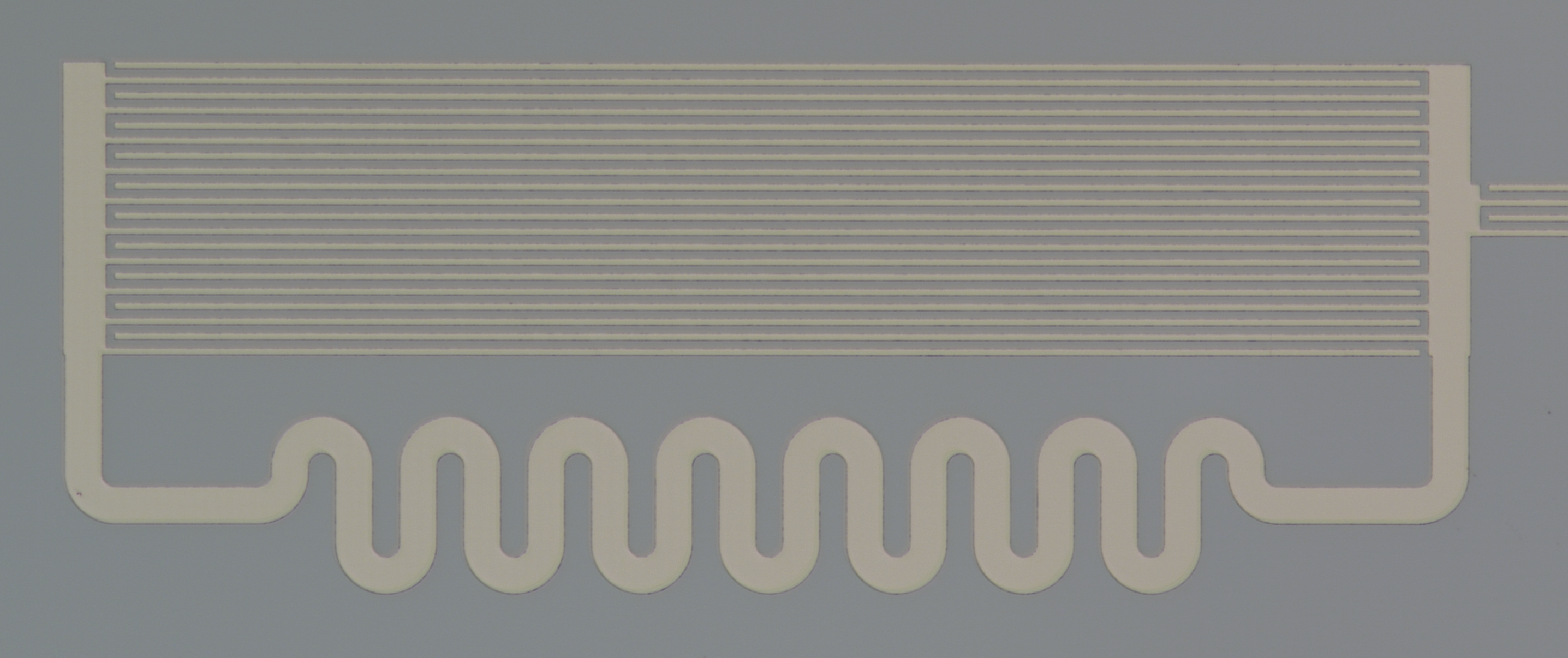}}}
     \put(0,-150){\color{black}\large{(b)}}
     \put(173, -153){\color{black}\linethickness{0.25mm}
     \frame{\includegraphics[scale=.05, angle=-90]{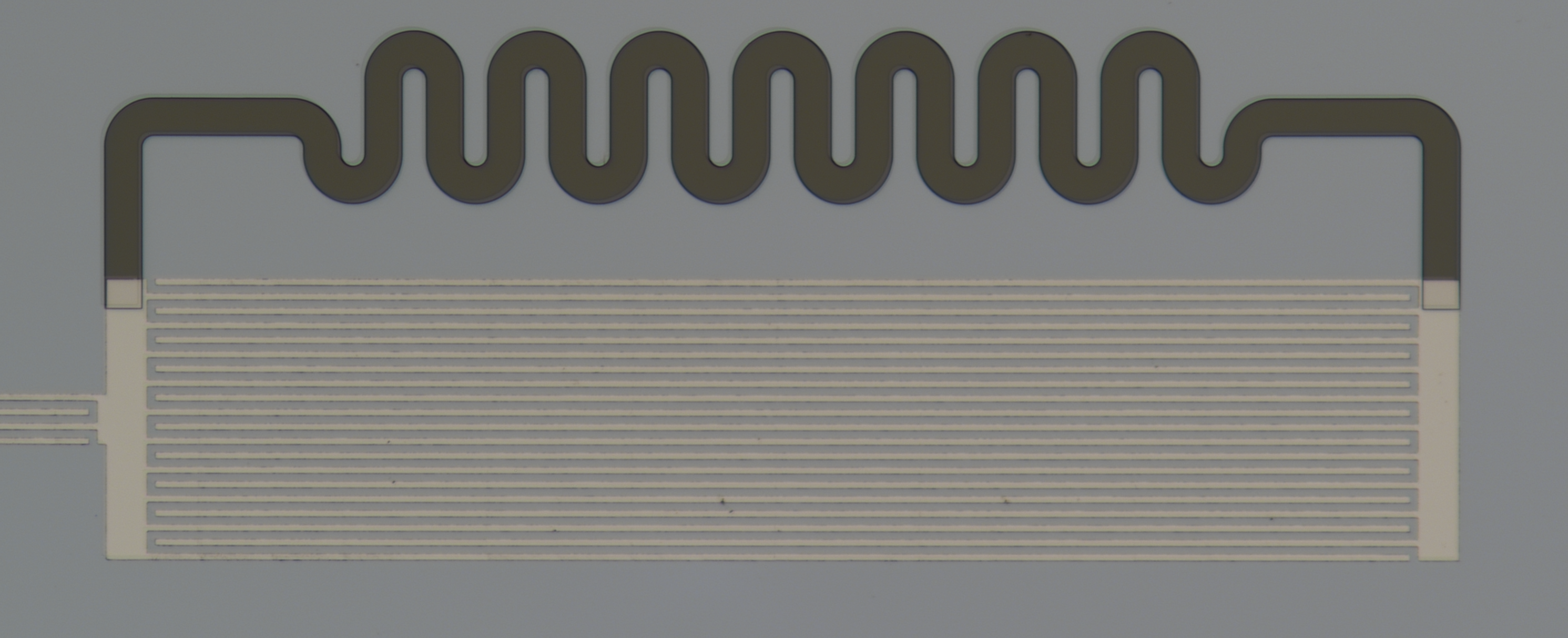}}}
     \put(173,-150){\color{black}\large{(c)}}
     \put(200,-20){\linethickness{0.35mm}\color{black}\polygon(2,0)(12,0)(12,12)(2,12)}
     \put(65, -76){\color{black}\linethickness{0.25mm}
     \frame{\includegraphics[scale=.12, angle=-90]{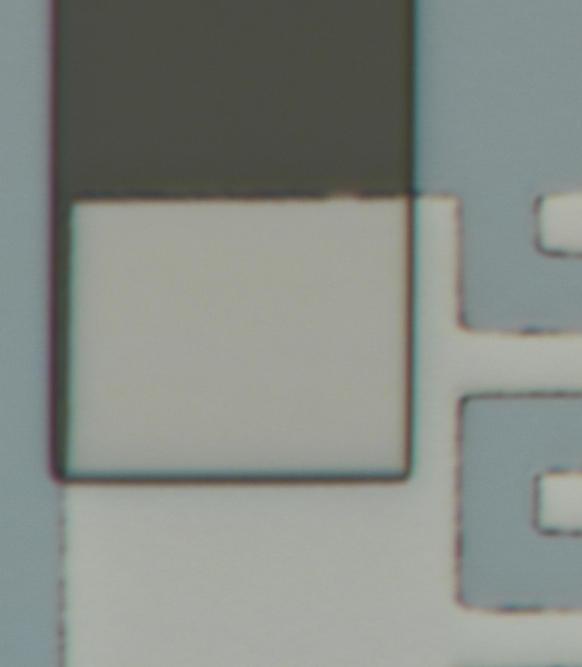}}}
     \put(78,-72){\color{black}\large{(d)}}

     \end{overpic}
     \vskip 150pt
     \caption{\footnotesize{(a), `alpha' resonator design. Top three resonators are fabricated with one layer of Nb. They have inductor width of 18, 20 and \SI{22}{\micro\meter} and everything else same. Bottom three resonators are same with top three except the inductors are all fabricated with TiN instead of Nb. (b), microscope image of all Nb resonator with inductor width of \SI{20}{\micro\meter} indicated in upper black square in (a). (c), microscope image of TiN resonator with inductor width of \SI{20}{\micro\meter} indicated in lower black square in (a). (d) End of the Nb interdigital capacitor and Nb/TiN interface as indicated in black square in (c). (e) Amplitude plot of one of the resonators}}
\label{fig::alphadesign}
\end{figure}
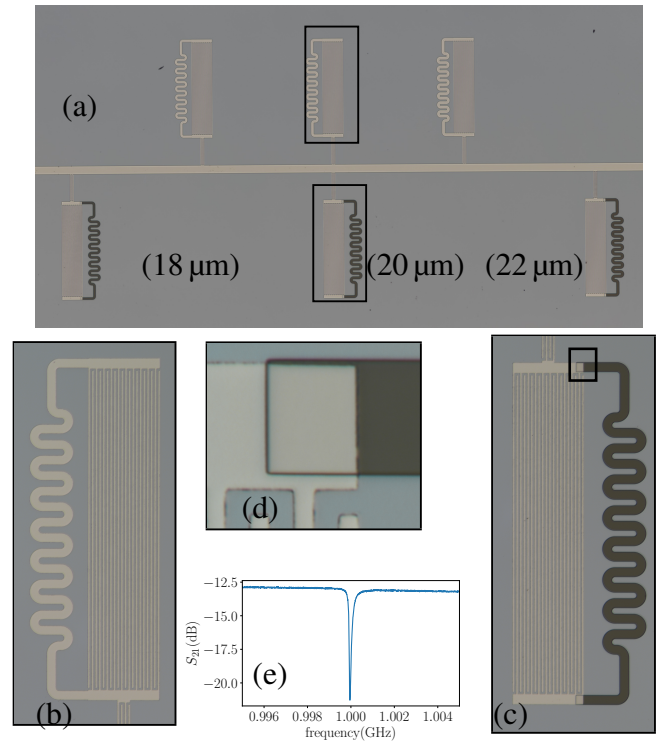
\begin{table}
\label{ta::alpha}
\caption{\label{tab:alphaparas} Parameters of the alpha design resonators. Nb stands for the upper three resonators in Figure \ref{fig::alphadesign} (a) and TiN are the lower three ones with TiN inductor. $w$ and $len$ are the design inductor width and length. $f_{o}$ and $Q_{o}$ are the measured resonator frequency and internal quality factor. $L_{k}$, $L_{g}$ and $C_{g}$ are the calculated kinetic inductance, geometric inductance and capacitance of the resonator in Section \ref{sec::Lkcal} \textbf{A}. }
\begin{ruledtabular}
\begin{tabular}{lccccccr}
\label{tab::alpha}
film & $w$ & $len$ & $f_{o}$ & $Q_{o}$ & $L_{g}$ & $C_{g}$ & $L_{k}$\\
 & (\SI{}{\um}) & (\SI{}{\um}) & (\SI{}{\GHz}) & $\times 10^{4}$ & (\SI{}{\nH}) & (\SI{}{\pF}) & (\SI{}{\nH})\\
\hline
    & 18 & 2600 & 5.90  & 5.53 & 1.42 & 0.58 & 0.0041\\
Nb   & 20 & 2600 & 6.03 & 5.75 & 1.32 & 0.57 & 0.004 \\
    & 22 & 2600 & 6.16 & 4.91 & 1.21 & 0.58 & 0.0039\\
\hline
    & 18 & 2600 & 0.92  & 4.14 & 1.42 & 0.58 & 51.5\\
TiN  & 20 & 2600 & 1.00 & 3.89 & 1.32 & 0.57 & 44.4 \\
    & 22 & 2600 & 1.10 & 4.35 & 1.21 & 0.58 & 35.8\\
\end{tabular}
\end{ruledtabular}
\end{table}

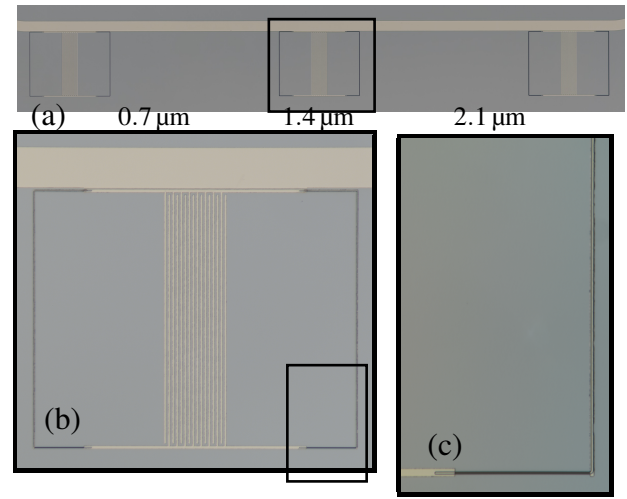
\begin{figure}
     \begin{overpic}[abs,unit=1pt,scale=.07, angle = -90, width=0.45\textwidth]{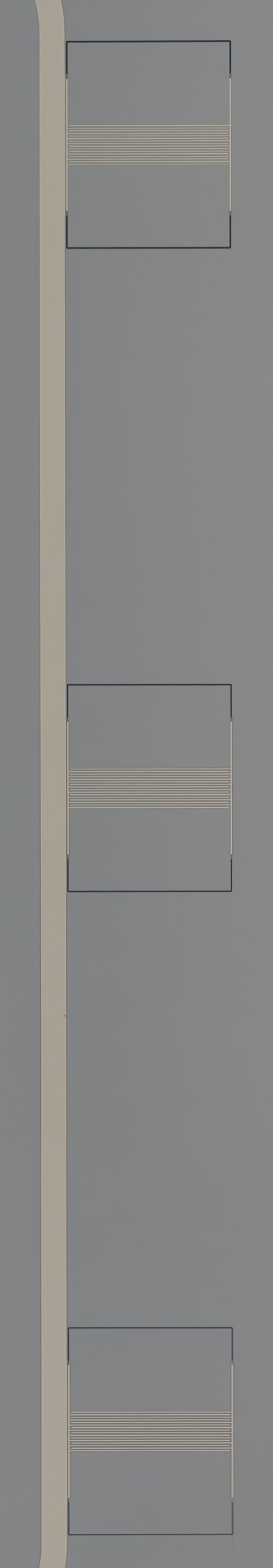}
     \put(95, 0){\linethickness{0.35mm}\color{black}\polygon(0,0)(40,0)(40,35)(0,35)}
     \put(38,-5){\color{black}\normalsize{\SI{0.7}{\micro\meter}}}
     \put(100,-5){\color{black}\normalsize{\SI{1.4}{\micro\meter}}}
     \put(165,-5){\color{black}\normalsize{\SI{2.1}{\micro\meter}}}
     \put(5,-5){\color{black}\large{(a)}}
     \put(0, -135){\color{black}\linethickness{0.6mm}
     \frame{\includegraphics[scale=.055, angle=90]{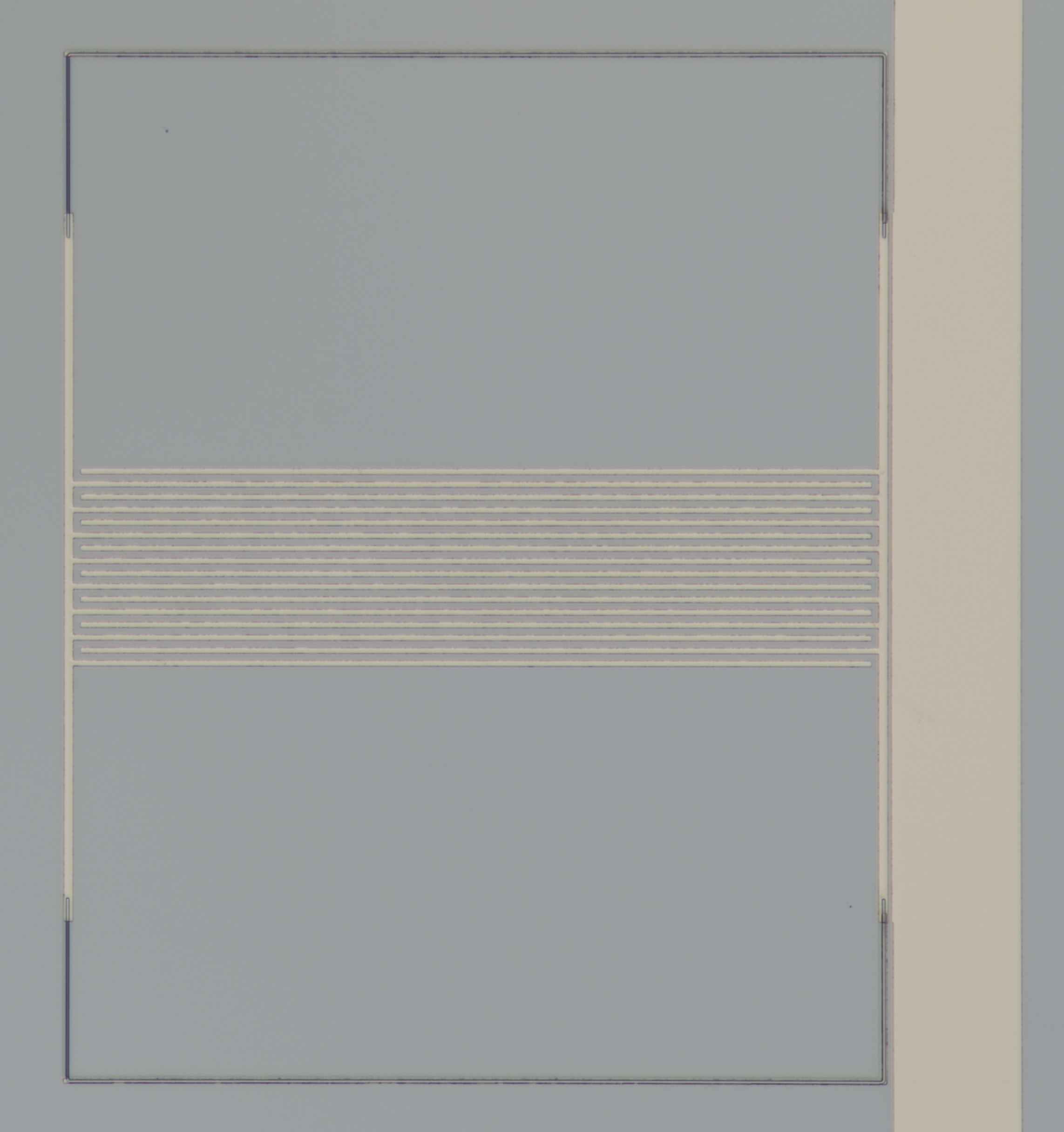}}}
     \put(10,-120){\color{black}\large{(b)}}
     \put(100,-140){\linethickness{0.35mm}\color{black}\polygon(2,0)(32,0)(32,44)(2,44)}
     \put(145, -144){\color{black}\linethickness{0.55mm}
     \frame{\includegraphics[scale=.03, angle=90]{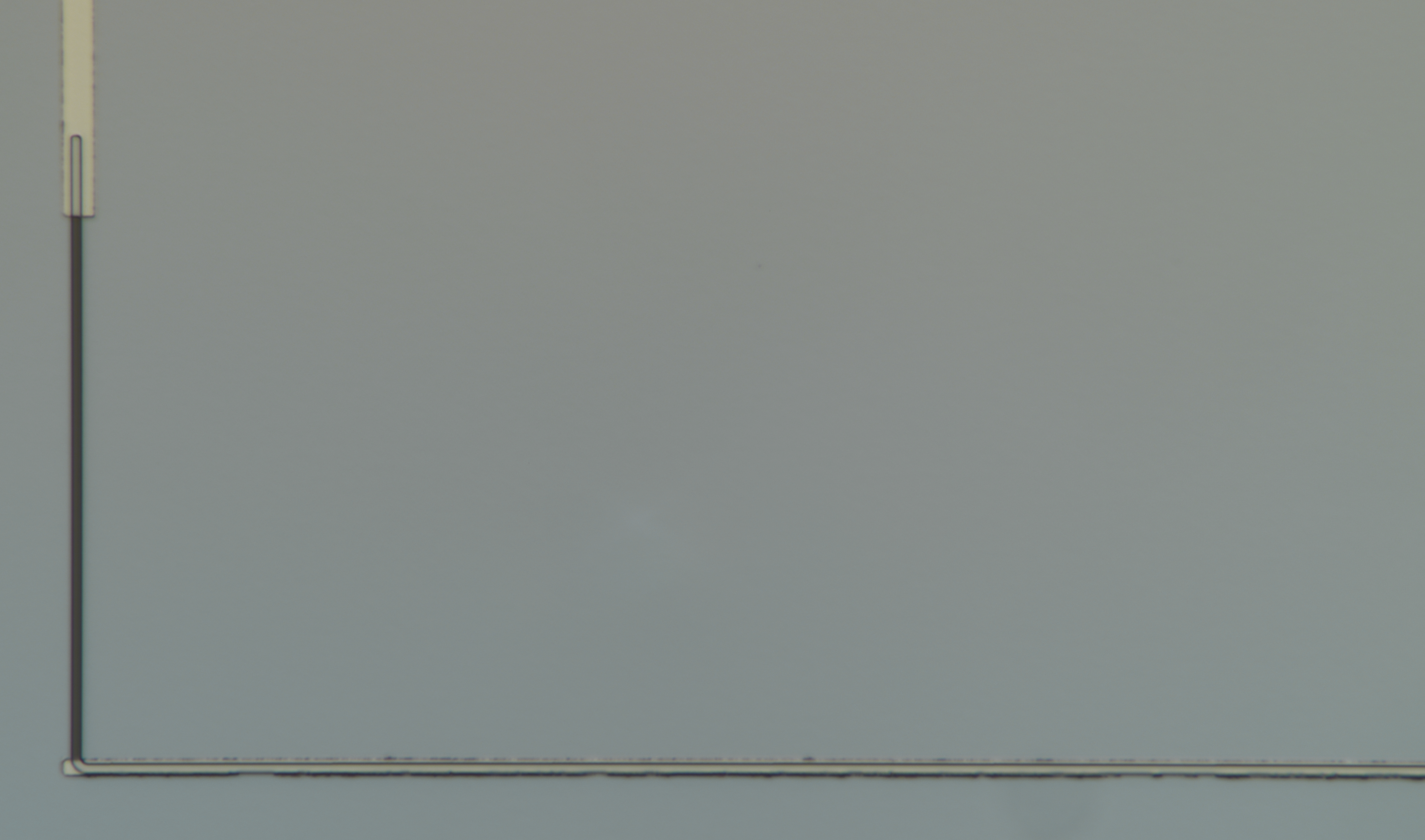}}}
     \put(155,-130){\color{black}\large{(c)}}
     \end{overpic}
     \vskip 160pt
     \caption{\footnotesize{Resonator design for flux coupled lumped element LC resonators. (b), optical microscope image of the resonator with inductor width of \SI{1.4}{\micro\meter} marked by the black square in (a). Light yellow colored film is Nb for readout and interdigital capacitor. Grey colored film is TiN for two inductors in parallel. (c) Image of the inductor joining the capacitor end marked by the gray square in (b).}}
\label{fig::fluxbiasdesign}
\end{figure}

The second group of resonators uses a design similar to the previously reported flux-coupled lumped-element LC resonator\cite{li2024} consisting of a closed superconducting loop with interdigitatal capacitor as shown in Figure \ref{fig::fluxbiasdesign}. The three resonators are patterned identically except for their inductor width of 0.7, 1.4 and \SI{2.1}{\micro\meter}. The inductors are fabricated with TiN while the interdigitated capacitors are fabricated with Nb. The measured resonant frequency $f_{o}$ and $Q_{o}$ are listed in Table \ref{tab:fb700para} in the row with inductor length `len' of `\SI{700}{\micro\meter}'.
\begin{table}
\caption{\label{tab:fb700para} Parameters of the flux coupled lumped element LC resonators. $w_{m}$ is averaged measured inductor width. All other parameters are same as in Table \ref{tab::alpha}.}
\begin{ruledtabular}
\begin{tabular}{lccccccr}
$len$ & $w$ & $w_{m}$ & $f_{o}$ & $Q_{o}$ & $L_{g}$ & $C_{g}$ & $L_{k}$\\
(\SI{}{\um}) & (\SI{}{\um}) & (\SI{}{\um}) & (\SI{}{GHz}) & $\times 10^{4}$ & (\SI{}{nH}) & (\SI{}{pF}) & (\SI{}{nH})\\
\hline
    & 0.7 & 0.71 & 0.77  & 5.34 & 1.59 & 0.227 & 378 \\
700 & 1.4 & 1.41 & 1.00 & 4.55  & 1.5 & 0.226 & 224\\
    & 2.1 & 2.11 & 1.15 & 4.64  & 1.36 & 0.244 & 156 \\
\hline
    & 0.7 & 0.71 & 2.17  & 4.47 & 1.41 & 0.24 & 44.6 \\
100 & 1.4 & 1.41 & 2.72 & 3.59  & 1.39 & 0.24 & 28.2 \\
    & 2.1 & 2.11 & 3.04 & 6.83 & 1.24 & 0.27 & 20.0 \\
\end{tabular}
\end{ruledtabular}
\end{table}

The third group of resonators have an additional Nb added to short the majority of the TiN inductor in the second group. The additional Nb reduces the TiN length in the group from \SI{700}{\micro\meter} 
to \SI{100}{\micro\meter}. The remainder of the inductor with the length of \SI{600}{\micro\meter} is covered with Nb film with width of \SI{4}{\micro\meter} and thickness of \SI{120}{\micro\meter}, shorting the high $L_{k}$ from the TiN film. The measured resonant frequencies (group \SI{100}{\micro\meter} in Table \ref{tab:fb700para}) are shifted to higher values as a result of the lower kinetic inductance in the modified pattern.
\section{Sheet kinetic inductance estimate}
\label{sec::Lkcal}
To fully characterize the sheet kinetic inductance $L_{k,\square}$ two methods were applied: calculating the total inductance of each resonator from its measured resonant frequency, and using a formula with the sheet resistance measured from DC samples. With the resonators the calculated total kinetic inductance was plotted as a function of the number of square ($\square$) using the geometry of each resonator. A linear fit to the nine data points yields a slope corresponding to the estimated $L_{k.\square}$. This calculated $L_{k,\square}$ is then compared to the simulated value using the resonators from the alpha design where we find consistency. In the second method $L_{k,\square}$ is directly calculated from the measured sheet resistance of the DC samples assuming TiN superconductor follows a BCS theory and found very close to the value obtained using the first method.
\subsection{Inductance simulations}
\label{sec::LkcalA}
Here we determine the sheet kinetic inductance as a function of inductor length/width through calculations of the geometric capacitance and measured resonance frequency. To calculate the geometric and kinetic parts of the impedance for all the resonators we simulate their resonant frequencies under two configurations: a frequency, $f_{o}$, with sheet kinetic inductance of $L_{k, \square} = 0$ and a second frequency, $f_{40}$, with $L_{k, \square} = 40~\textrm{pH}/\square$.
Each of the resonant frequencies are defined through their lumped element components as in the equations below:
\begin{eqnarray}
    f_{o} &=& \frac{1}{2\pi \sqrt{L_{g}C_{g}}} \\
    f_{40} &=& \frac{1}{2\pi \sqrt{(L_{g}+L_{40})C_{g}}}
\end{eqnarray}
and their frequency ratio gives: 
\begin{eqnarray}
       \label{eq::foratio}
         \left(\frac{f_{o}}{f_{40}}\right)^{2} &=& \frac{L_{g}+L_{40}}{L_{g}}
\end{eqnarray}
The total kinetic inductance is fixed by the assumed sheet kinetic inductance and inductor geometry, $L_{40} = 40\times\text{length/width}$. The geometric inductance is then calculated through the two resonance ratio (Equation \ref{eq::foratio}) $L_{g} = L_{40}/((f_{o}/f_{40})^{2}-1)$. The geometric capacitance is calculated from the zero kinetic inductance resonance $C_{g} = 1/(L_{g}(2\pi f_{o})^{2})$.
With the geometric inductance and capacitance derived from our simulations using the above technique, the kinetic inductance $L_{k,m}$ at a specific sheet inductance $m$ with resonant frequency $f_{m}$ can then be calculated through equation below:
\begin{eqnarray}
\label{eqn::Lkcal}
    L_{k,m} &=& \frac{1}{C_{g}(2\pi f_{m})^{2}}
\end{eqnarray}
where we have ignored the geometric inductance $L_{g}$ for $L_{g}\ll L_{k,m}$.
\begin{figure}
     \begin{overpic}[abs,unit=1pt,scale=.05,width=0.45\textwidth]{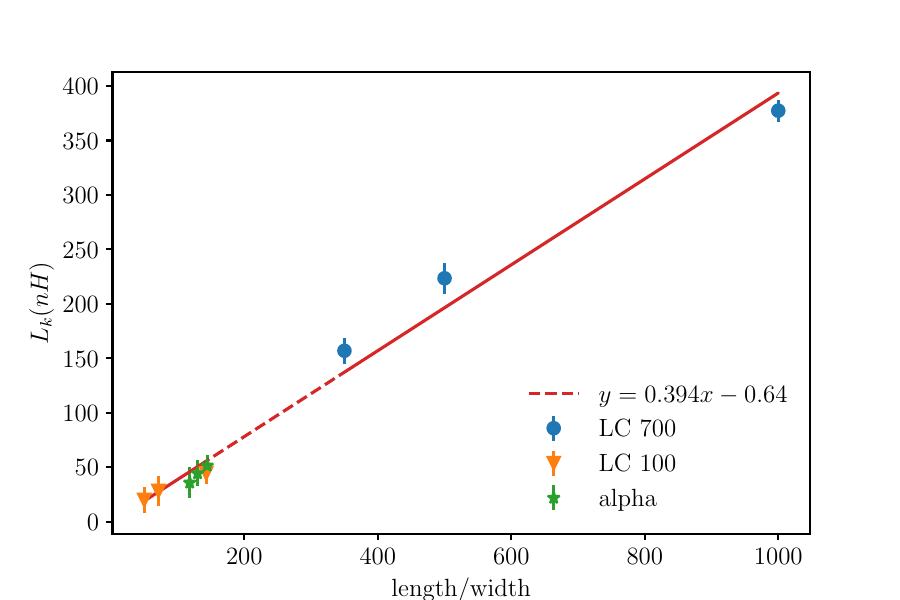}
    \put(60,110){\color{black}\large{$L_{k,\square}=394~\textrm{pH}/\square$}}
     \end{overpic}
     \vskip 3pt
     \caption{\footnotesize{Fitted $L_{k,\square}$ for all three group of resonators. LC 700 are the original three flux biased LC resonators. Alpha stands for the three resonators in the alpha design with TiN inductor. LC 100 are the three resonators by shorting \SI{600}{\micro\meter} lenght of the TiN inductor with Nb microstrip on top. The total inductor is the left \SI{100}{\micro\meter} TiN. Straight line is fitting a linear function to all the data points. Error bar in each data point is calculated from the uncertainty of fitting $S_{21}$ equation to the measured resonance curves\cite{carter2017}.}}
\label{fig::lkcal3}
\end{figure}

The total inductance from our measurements of all nine TiN resonators is calculated using equation (\ref{eqn::Lkcal}) and plotted in figure \ref{fig::lkcal3} as a function of the TiN length divided by its width. After fitting to a line the slope of the equation provides our estimate of the sheet kinetic inductance $L_{k, \square} = 394~\textrm{pH}/\square$.

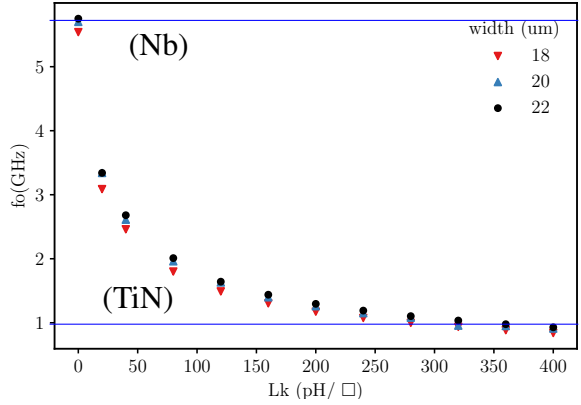
\begin{figure}
     \begin{overpic}[abs,unit=1pt,scale=.8,width=0.5\textwidth]{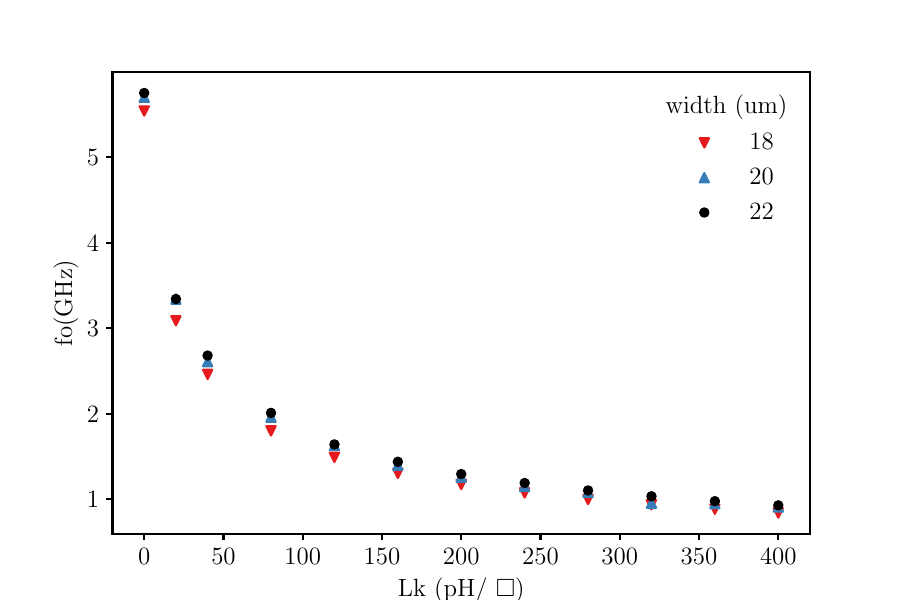}
     \put(60,130){\color{black}\large{(Nb)}}
     \put(50,34){\color{black}\large{(TiN)}}
     \put(30,143){\color{blue}\line(1,0){200}}
     \put(30,28){\color{blue}\line(1,0){200}}
     \end{overpic}
     \vskip 3pt
     \caption{\footnotesize{Simulated resonant frequency as function of $L_k$ for alpha design resonators. Label `(Nb)' and `(TiN)' mark the resonance frequencies match with the all Nb resonators and the resonators with TiN inductor respectively.}}
\label{fig::sonnetcurve}
\end{figure}

As a consistency check of our estimated $L_{k, \square}$, we simulated the three resonators for the alpha design where we vary the sheet $L_{k,\square}$  from 0 to $400~\textrm{pH}/\square$ in steps of $20~\textrm{pH}/\square$. Fig \ref{fig::sonnetcurve} shows the simulated $f_{o}$ as function of $L_{k,\square}$. The simulated $f_{o}$ for $L_{k,\square}=0$ (blue line with label `(Nb)') matches the measured resonant frequency of the all Nb resonators which has negligible kinetic inductance. 
At high  values of $L_{k, \square}$, the dependence of $f_{o}$ on $L_{k,\square}$ is weak, with values of $\sim400~\textrm{pH}/\square$ showing resonant frequencies that are consistent with what we measured. 
\subsection{Estimating kinetic inductance from measured resistance}
Assuming TiN follows conventional BCS behavior, the sheet inductance $L_{k,\square}$ can also be calculated from its sheet resistance with equation $L_{k, \square} = \hbar R_{\square}/\pi\Delta_{0}$, where the sheet resistance $R_{\square}=R\times\text{Width}/\text{Length}$ is measured just above Tc and $\Delta_{0} = 1.76~k_{B}T_{c}$ is the superconducting energy gap for TiN \cite{escoffier2004,pracht2012}. For our sputtered films, the measured $R_{\square} = 475~\Omega/\square$, which corresponds to a sheet inductance of $L_{k, \square}=423~\textrm{pH}/\square$. This is close to the value estimated from our first method (\ref{sec::LkcalA}).
\section{Structural and Morphological Characterization}
\begin{figure}
     \begin{overpic}[abs,unit=1pt,scale=0.55,width=0.45\textwidth]{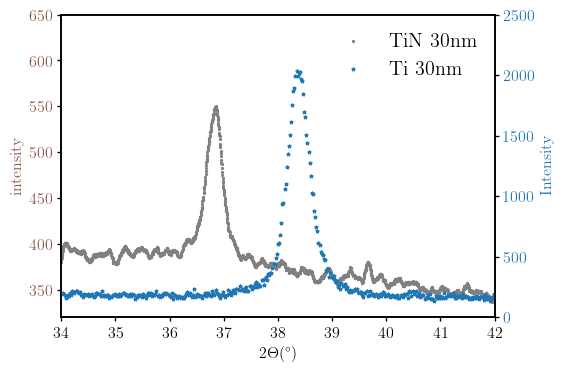}
     \put(60,110){\color{black}\normalsize{\sffamily (111)}}
     \put(95,130){\color{black}\normalsize{\sffamily (002)}}
     \end{overpic}
     \vskip 3pt
     \caption{\footnotesize{XRD image of unpatterned TiN (brown curve) compared to Ti (blue curve) films with 30nm thickness.}}
\label{fig::xrd26nm}
\end{figure}

The structural and morphological properties of the films were investigated using X-ray diffraction (XRD) and atomic force microscopy (AFM). Figure \ref{fig::xrd26nm} presents the XRD patterns of the TiN (grey curve) and Ti (blue curve) films. For the TiN layer, a single diffraction peak is observed at $2\Theta = 36.8^{\circ}$, corresponding to the (111) reflection of face-centered cubic TiN. The measured full width at half maximum (FWHM) of this peak is $0.3637^{\circ}$. Absence of additional reflections such as (200) and (220) indicates a strong (111) preferred orientation, suggesting that the film is polycrystalline with distinct texture. This observation is consistent with previously reported results for TiN films deposited by sputtering\cite{vissers2010, ohya2013, jaim2015}. The Ti film exhibits a peak at $2\Theta = 38.3^{\circ}$, corresponding to the (002) reflection plane, revealing a hexagonal close-packed structure with strong texture. Notably, the TiN film does not display any residual Ti-related peaks, indicating that the titanium has reacted completely with nitrogen during the deposition to form the desired (111) TiN.

\begin{figure}
     \begin{overpic}[abs,unit=1pt,scale=0.55,width=0.45\textwidth]{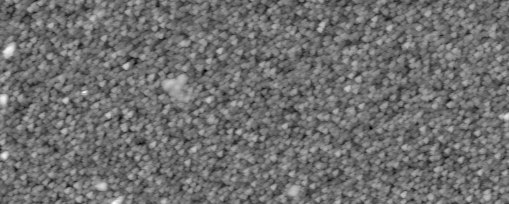}
     \put(188,8){\color{white}\rule{20pt}{2pt}} 
     \put(184,15){\color{white}\normalsize{\SI{200}{nm}}}
     \end{overpic}
     \vskip 3pt
     \caption{\footnotesize{AFM scan image of unpatterned TiN showing the grain size and surface roughness of the film. The image has been enhanced through FFT (fast Fourier transformation) and filtering out lower frequency pixels. Extremely larger grains are surface contamination.}}
\label{fig::AFM26nm}
\end{figure}

Average crystallite size of the TiN film was estimated using the Scherrer equation\cite{scherrer1918}, based on the broadening of the observed peak. The calculated crystallite size was approximately \SI{24.4}{nm}. The \SI{27}{nm} thickness assumes that the film is formed by vertically aligned nanocrystalline grains extending from the substrate to the surface. This interpretation is supported by AFM surface topography (Figure \ref{fig::AFM26nm}), which reveals well-defined grain structures consistent with (111)-oriented columnar growth. The average RMS surface roughness was \SI{1.62}{nm}.

\section{conclusion}
In conclusion, we have fabricated demonstrated TiN thin film with very high sheet kinetic inductance $L_{k,\square} \approx 400~\textrm{pH}/\square$. This film could find broad application in quantum circuits, parametric amplifiers and superconducting qubits. 
Further investigation of this film will be extensive study of the film properties, including loss and sheet impedance, as a function of $N_{2}$ flow rate and sputtering power as well as film thickness.
\begin{table}
\caption{\label{tab:paras} properties of reactively sputtered TiN film. The thickness (t) of the film is measured via surface profilometer. The critical temperature (Tc) shown is the temperature at a $50\%$ reduction in resistivity ($\rho$) from \SI{3.8}{K}. $Q_{o}$ is the average value of all the measured resonators. Sheet resistance $R_{s, 300}$~($\Omega/\square$) is four probe measurement with the DC sample. Sheet kinetic inductance values are measured with the method in section \ref{sec::Lkcal}.}
\begin{ruledtabular}
\begin{tabular}{l|c}
 parameters: & values \\
\hline
$Q_{o}$ & $10^{4}$   \\
$T_{c}$~(K) & 3.8  \\
$R_{s, 300}$~($\Omega/\square$) & 475  \\
t~(nm) & 27  \\
$L_{k, \square}~(\textrm{pH}/\square)$ & 394  \\
\end{tabular}
\end{ruledtabular}
\end{table}
\section*{Data Availability Statement}
Data used in this work is available on reasonable request.
\vspace*{-0.3cm}
\section*{Acknowledgement}
Work at the University of Chicago Pritzker Nano-fabrication Facility was supported by NASA grant 80NSSC22K174. Work at Argonne National Lab, including work performed at the Center for Nanoscale Materials, a U.S. Department of Energy Office of Science User Facility, is supported by the U.S. Department of Energy, Office of Science, Office of High Energy Physics and Office of Basic Energy Sciences, under Contract No. DE-AC02-06CH11357. This material is based upon work supported by the U.S. Department of Energy Office of Science National Quantum Information Science Research Centers. The work at Q-NEXT includes concept development, design, fabrication, testing, and modeling of devices.
\vspace*{-0.3cm}
\appendix
\section{reactive sputtering process parameters}
\label{app::dfdphi}
\begin{figure}[h]
     \begin{overpic}[abs,unit=1pt,scale=.05,width=0.215\textwidth]
     {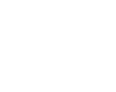}
     \put(-77, -30){\color{white}\linethickness{0.55mm}
     \frame{\includegraphics[scale=.9, angle=0]{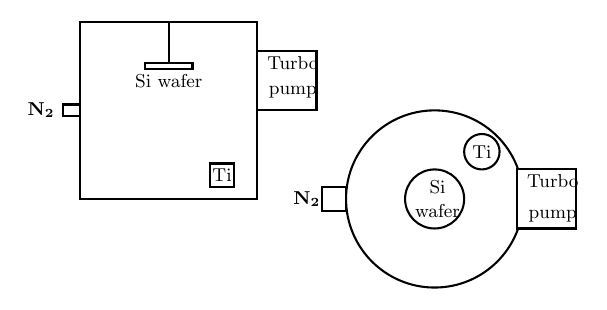}}}
     \put(-20,5){\color{black}\large{(a)}}
     \put(100,65){\color{black}\large{(b)}}
     \end{overpic}
     \vskip 10pt
     \caption{\footnotesize{(a), side view of the sputtering chamber showing the locations of titanium target, turbo pumping and \ce{N2} gas inlet. (b), top view.}}
\label{fig::chamber}
\end{figure}

\begin{table}
\caption{\label{tab:paras} parameters of TiN sputtering process.}
\begin{ruledtabular}
\begin{tabular}{lc}
 \textbf{Parameter} & \textbf{Value}\\
\hline
machine model & AJA ATC 2200 \\
$N_{2}$ flow rate~(sccm) & 3.6 \\
Chamber pressure~(mTorr) & 3  \\
$Ar$ flow rate~(sccm) & 30 \\
deposition Rate~($\dot{A}$/min) & 9\\
DC Power~(W) & 200 \\
target and substrate distance~(cm) & 12 \\
magnet configuration & unbalanced
\end{tabular}
\end{ruledtabular}
\label{tab::spupara}
\end{table}

\section{DC sample and measurement setup}
\label{app::DCRFsetup}
\begin{figure}[h]
     \begin{overpic}[abs,unit=1pt,scale=.05,width=0.215\textwidth]
     {white.png}
     \put(-67, -30){\color{white}\linethickness{0.55mm}
     \frame{\includegraphics[scale=.04, angle=0]{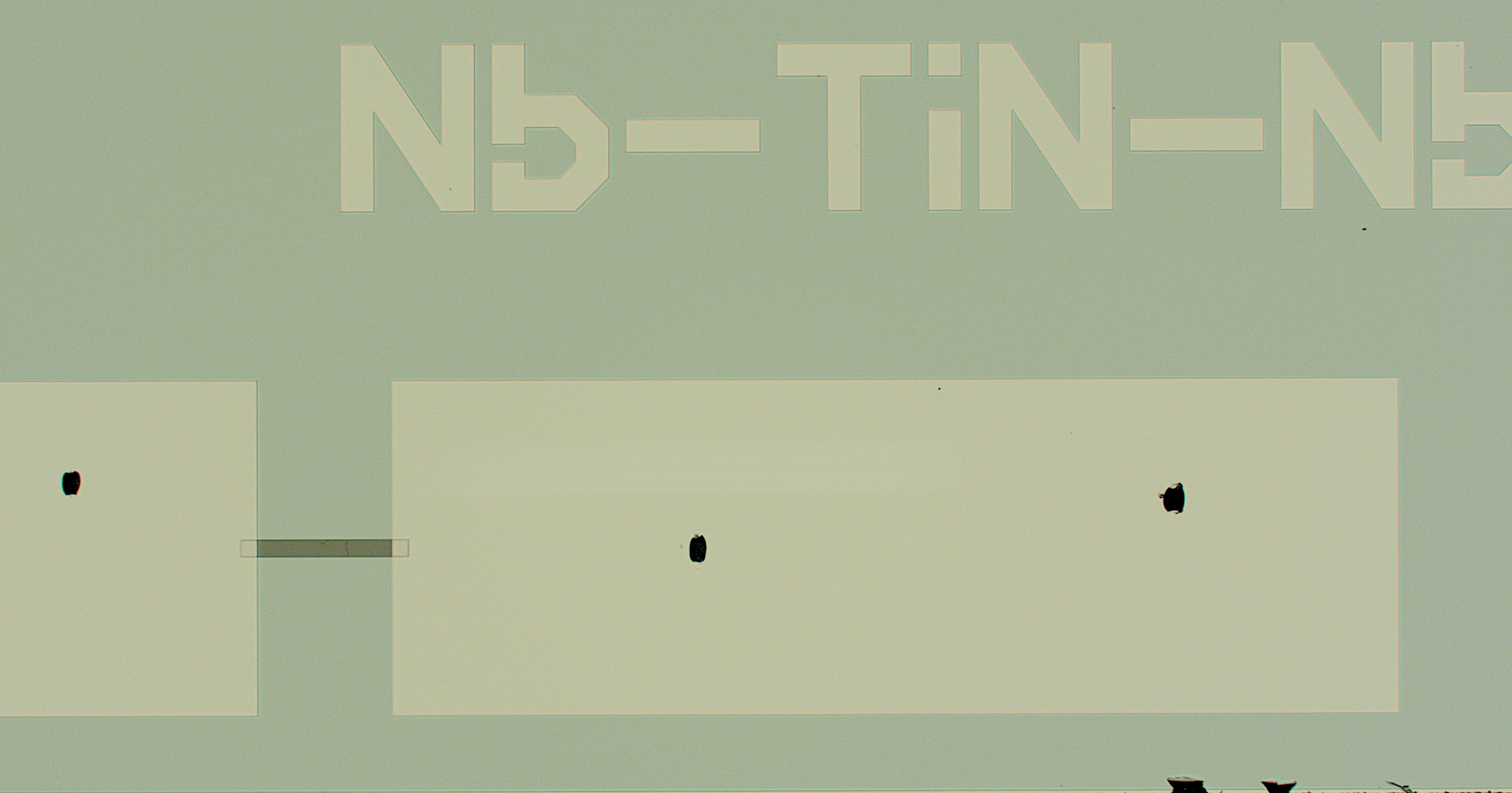}}}
     \put(-65,12){\color{black}\large{(a)}}
      \put(-52,-16){\linethickness{0.25mm}\color{blue}\polygon(2,0)(12,0)(12,6)(2,6)}
     \put(-67, 35){\color{blue}\linethickness{0.55mm}
     \frame{\includegraphics[scale=.04, angle=0]{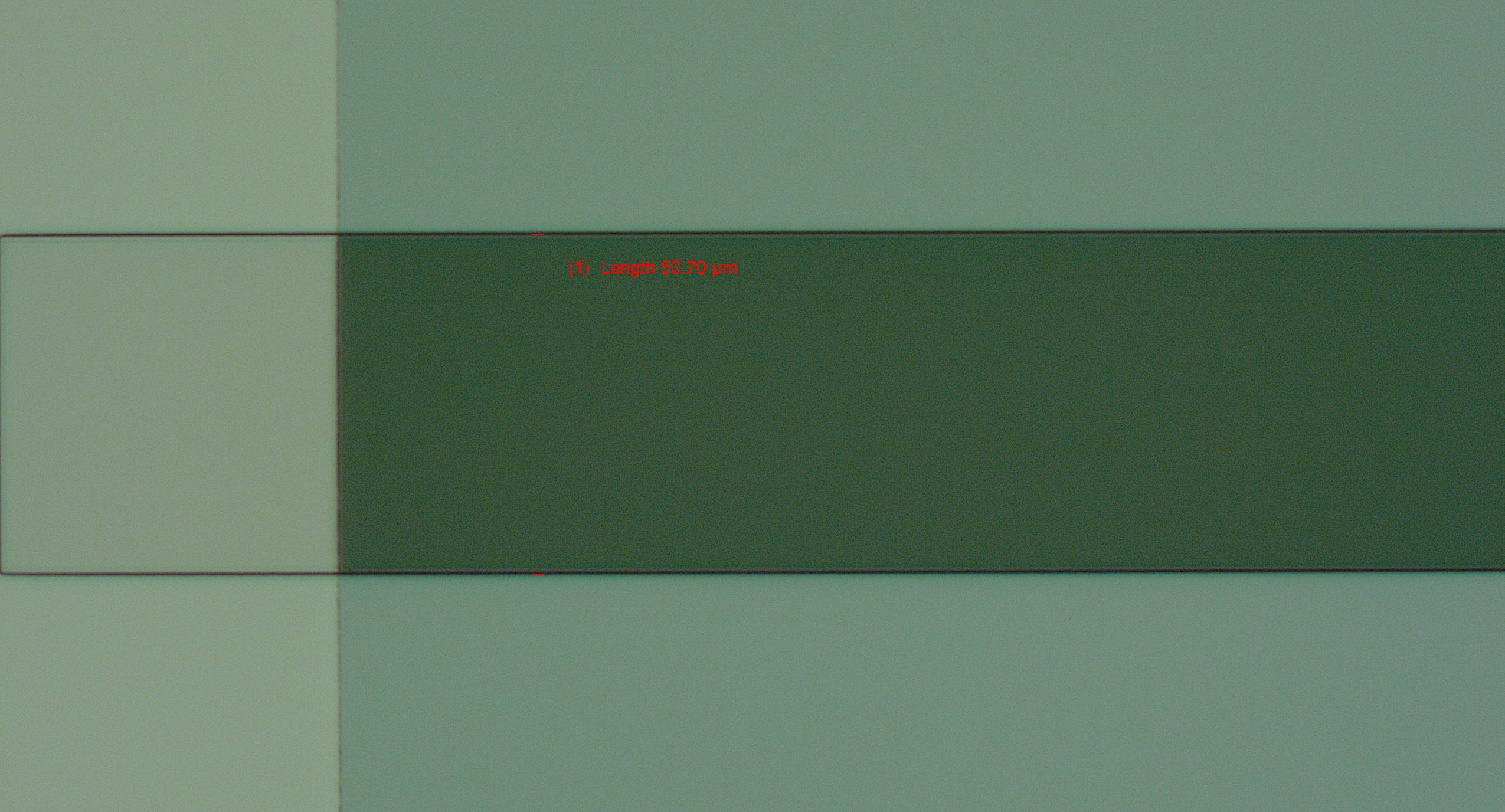}}}
     \put(-65, 41){\color{black}\large{(b)}}
     \put(40, -30){\color{white}\linethickness{0.55mm}
     \frame{\includegraphics[scale=.35, angle=0]{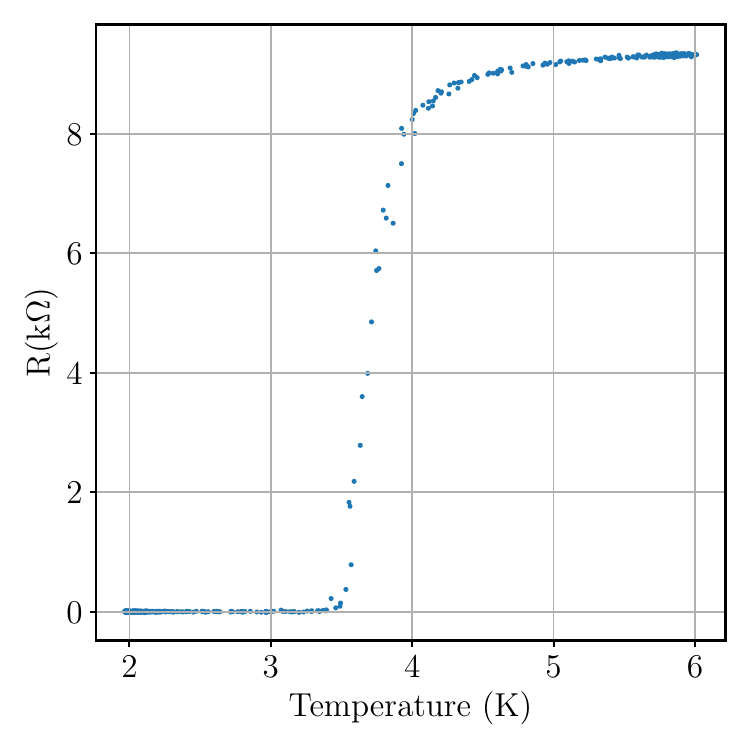}}}
     \put(65,77){\color{black}\large{(c)}}
     \end{overpic}
     \vskip 20pt
     \caption{\footnotesize{(a), optical microscopy images of dog bone style DC samples for Tc measurement. The bigger pads are for wire bonding and the resistance is dominated by the bridge in the middle with dimension of \SI{50}{\micro\meter}$\times$\SI{400}{\micro\meter}. (b), zoom in to the blue square in (a) shows the narrow TiN strip. (c) Resistance transition curve of TiN bridge DC samples as function of temperature. Tc is the temperature value when its resistance dropped to half of its resistance before transition starts.}}
\label{fig::dcsample}
\end{figure}
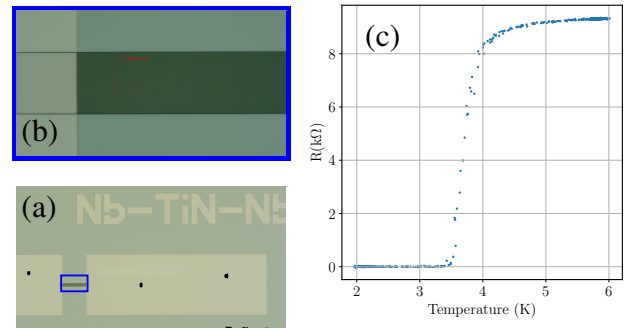

Figure \ref{fig::dcsample} (a) shows the design of the DC samples used for $T_{c}$ measurement. The center bridge with dimension of \SI{50}{\micro\meter}$\times$\SI{400}{\micro\meter} is made of TiN. The two contact pads are for wire bonding and are made of both TiN and Nb. Both design show close Tc values. Sample resistance is measured with four wire measurement with $I^{+}, V^{+}$ and  on one pad and $I^{-}, V^{-}$ on the other pad. Kiethly 6221 current source is used for current bias. Measured resistance transition curve is shown in (b).

\begin{table}[htbp]
\caption{\label{tab:paras} Properties of sputtered TiN film. The thickness ($t$) of the film is measured via surface profilometer. The critical temperature ($T_c$) shown is the temperature at a $50\%$ reduction in resistivity ($\rho$) from \SI{3.8}{K}. $Q_{o}$ is the average value of all the measured resonators. Sheet resistance $R_{s, 300}$~($\Omega/\square$) is a four-probe measurement with the DC sample. Sheet kinetic inductance values are measured using the method in Section~\ref{sec::Lkcal}.}
\begin{tabularx}{0.5\linewidth}{|>{\raggedright\arraybackslash}X| >{\raggedleft\arraybackslash}X|}
\toprule
\textbf{Parameter} & \textbf{Value} \\
\midrule
$Q_{o}$ & $10^{4}$ \\
$T_{c}$~(K) & 3.8 \\
$R_{s, 300}$~($\Omega/\square$) & 475 \\
$t$~(nm) & 27 \\
$L_{k, \square}$~(pH$/\square$) & 394 \\
\bottomrule
\end{tabularx}
\end{table}

\bibliography{reff}

\end{document}